\documentclass[fleqn,usenatbib,useAMS]{mnras}

\usepackage{graphicx}	
\usepackage{amsmath}	
\usepackage{amssymb}	
\usepackage{multicol}        
\usepackage{bm}		
\usepackage{pdflscape}	
\usepackage{threeparttable}

\newcommand{\kms}{\,km\,s$^{-1}$} 
\newcommand{\ms}{\,m\,s$^{-1}$} 

\newcommand{\ktwo}{\emph{K2}} 

\newcommand{\target}{K2-100}
\newcommand{\pyan}{\texttt{pyaneti}}

\newcommand{\logr}{$\log R'_{\rm HK}$}

\newcommand\vsini{$v$\,sin\,$i_\star$}    
 
\newcommand\vmic{$v_{\rm mic}$}
\newcommand\vmac{$v_{\rm mac}$}

\newcommand\teff{$T_{\rm eff}$}
\newcommand\logg{log\,{\it g$_\star$}}

\newcommand{\smass}[1][$M_{\odot}$]{ $ 1.15 \pm 0.05 $ #1} 
\newcommand{\sradius}[1][$R_{\odot}$]{ $1.24\pm 0.05 $ #1}
\newcommand{\stemp}[1][$\mathrm{K}$]{ $ 5945 \pm110 $ #1 }
\newcommand{\Tzerob}[1][days]   {$7140.71941 \pm 0.00027 $~#1} 
\newcommand{\Pb}[1][days]   {$1.6739035 \pm 0.0000004$~#1} 
\newcommand{\bb}[1][ ]   {$0.791 \pm 0.014 $~#1} 
\newcommand{\arb}[1][ ]   {$5.21 \pm 0.13 $~#1} 
\newcommand{\rrbKtwo}[1][ ]   {$0.02867 \pm 0.00028 $~#1} 
\newcommand{\kb}[1][${\rm m\,s^{-1}}$]   {$10.6 \pm 3.0 $~#1} 
\newcommand{\mpb}[1][$M_{\oplus}$]   {$21.8 \pm 6.2 $~#1} 
\newcommand{\rpbKtwo}[1][$R_{\oplus}$]   {$3.88 \pm 0.16 $~#1} 

\newcommand{\ib}[1][deg]   {$81.27 \pm 0.37 $~#1} 
\newcommand{\ab}[1][AU]   {$0.0301 \pm 0.0014 $~#1} 
\newcommand{\insolationb}[1][${\rm F_{\oplus}}$]   {$1915 _{ - 165 } ^ { + 178}$~#1} 
 
\newcommand{\densspb}[1][${\rm g\,cm^{-3}}$]   {$0.85 _{ - 0.10 } ^ { + 0.12 }$~#1} 
\newcommand{\denpb}[1][${\rm g\,cm^{-3}}$]   {$2.04 _{ - 0.61 } ^ { + 0.66 }$~#1} 
\newcommand{\grapb}[1][${\rm cm\,s^{-2}}$]   {$1536 _{ - 442 } ^ { + 436}$~#1} 
\newcommand{\grapparsb}[1][${\rm cm\,s^{-2}}$]   {$1421 _{ - 413 } ^ { + 427}$~#1} 
\newcommand{\Teqb}[1][K]   {$1841 \pm 41$~#1} 
\newcommand{\qoneKtwo}[1][]   {$0.27 _{ - 0.07 } ^ { + 0.08 }$~#1} 
\newcommand{\qtwoKtwo}[1][]   {$0.13 _{ - 0.10 } ^ { + 0.19 }$~#1}

\newcommand{\qoneG}[1][]   {$0.03 _{ - 0.02 } ^ { + 0.06 }$~#1} 
\newcommand{\qtwoG}[1][]   {$0.40 _{ - 0.28 } ^ { + 0.36 }$~#1}

\newcommand{\qonebandone}[1][]   {$0.73 _{ - 0.26 } ^ { + 0.19 }$~#1} 
\newcommand{\qtwobandone}[1][]   {$0.49 _{ - 0.15 } ^ { + 0.13 }$~#1}

\newcommand{\qonebandtwo}[1][]   {$0.57 _{ - 0.26 } ^ { + 0.27 }$~#1} 
\newcommand{\qtwobandtwo}[1][]   {$0.47 _{ - 0.23 } ^ { + 0.22 }$~#1}

\newcommand{\HARPSN}[1][${\rm m\,s^{-1}}$]   {$34.393 \pm 0.003 $~#1} 
\newcommand{\RHK}[1][${\rm m\,s^{-1}}$]   {$-4.45 \pm 0.01 $~#1} 
\newcommand{\BIS}[1][${\rm m\,s^{-1}}$]   {$-0.04 \pm 0.04 $~#1} 
\newcommand{\jHARPSN}[1][${\rm m\,s^{-1}}$]   {$2.60 _{ - 2.05 } ^ { + 3.15 }$~#1} 
\newcommand{\jRHK}[1][${\rm m\,s^{-1}}$]   {$0.0030 \pm 0.0021 $~#1} 
\newcommand{\jBIS}[1][${\rm m\,s^{-1}}$]   {$291 _{ - 24 } ^ { + 27 }$~#1} 
\newcommand{\jtrKtwo}[1][]   {$40 \pm 4 $~#1} 
\newcommand{\jtrG}[1][]   {$ 267 \pm 50 $~#1} 
\newcommand{\jtrSC}[1][]   {$52 \pm 4 $~#1} 
\newcommand{\jtrbandone}[1][]   {$ 1321 \pm 27$~#1} 
\newcommand{\jtrbandtwo}[1][]   {$1419  \pm 30$~#1} 
\newcommand{\jtrbandthree}[1][]   {$1919\pm 38$~#1} 
\newcommand{\jArvc}[1][]   {$0.0058 _{ - 0.0037 } ^ { + 0.0049 }$~#1} 
\newcommand{\jArvr}[1][]   {$0.0421 _{ - 0.0095 } ^ { + 0.0147 }$~#1} 
\newcommand{\jAhk}[1][]   {$0.0242 _{ - 0.0055 } ^ { + 0.0079 }$~#1} 
\newcommand{\jAbisc}[1][]   {$0.020 _{ - 0.059 } ^ { + 0.061 }$~#1} 
\newcommand{\jAbisr}[1][]   {$-0.086 _{ - 0.049 } ^ { + 0.037 }$~#1} 
\newcommand{\jle}[1][]   {$31.2 _{ - 6.3 } ^ { + 7.6 }$~#1} 
\newcommand{\jlp}[1][]   {$0.558 _{ - 0.069 } ^ { + 0.082 }$~#1} 
\newcommand{\jPGP}[1][]   {$4.315 \pm 0.014 $~#1} 
\newcommand{\rrbG}[1][ ]   {$0.0308 \pm 0.0011 $~#1} 
\newcommand{\rrbkepler}[1][ ]   {$0.0286 \pm 0.0003 $~#1} 
\newcommand{\rrbbandr}[1][ ]{$0.0241 \pm 0.0015 $~#1} 
\newcommand{\rrbbandi}[1][ ]   {$0.0263 \pm 0.0015 $~#1} 
\newcommand{\rrbbandz}[1][ ]   {$0.0281 \pm 0.0019  $~#1} 
 
\defcitealias{Rajpaul2015}{R15}
\defcitealias{Mann2017}{M17}
\defcitealias{kubyshkina2018}{K18b}
\defcitealias{kubyshkina2019}{K19}

\usepackage[T1]{fontenc}
\usepackage{ae,aecompl}

\usepackage{newtxtext,newtxmath}

\title[RV confirmation of K2-100b]{
Radial velocity confirmation of K2-100b: a young, highly irradiated, and low density transiting hot Neptune}

\author[Barrag\'an et al.]
{O.~Barrag\'an$^{1,}$\thanks{Contact e-mail: \href{mailto:oscar.barraganvillanueva@physics.ox.ac.uk}{oscar.barraganvillanueva@physics.ox.ac.uk}},
S.~Aigrain$^{1}$,
D.~Kubyshkina$^{2}$,
D.~Gandolfi$^{3}$,
J.~Livingston$^{4}$,
\newauthor
M.~C.~V.~Fridlund$^{5,6}$, 
L.~Fossati$^{2}$,
J.~Korth$^{7}$,
H.~Parviainen$^{8,9}$,
L.~Malavolta$^{10}$,
E.~Palle$^{8,9}$,
\newauthor
H.~J.~Deeg$^{8,9}$,
G.~Nowak$^{8,9}$,
V.~M.~Rajpaul$^{11}$,
N.~Zicher$^{1}$,
G.~Antoniciello$^{12}$,
N.~Narita$^{13,14,15,8}$,
\newauthor
S.~Albrecht$^{16}$,
L.~R.~Bedin$^{17}$,
J.~Cabrera$^{18}$,
W.~D.~Cochran$^{19}$,
J.~de~Leon$^{4}$,
Ph.~Eigm\"{u}ller$^{18}$,
\newauthor
A.~Fukui$^{20}$,
V.~Granata$^{12}$,
S.~Grziwa$^{7}$,
E.~Guenther$^{21}$,
A.~P.~Hatzes$^{21}$,
N.~Kusakabe$^{13,15}$,
\newauthor
D.~W.~Latham$^{22}$,
M.~Libralato$^{23}$,
R.~Luque$^{8,9}$,
P.~Monta\~n\'es-Rodr\'iguez$^{8,9}$, 
F.~Murgas$^{8,9}$,
\newauthor
D.~Nardiello$^{12}$,
I.~Pagano$^{10}$,
G.~Piotto$^{12}$,
C.~M.~Persson$^{5,6}$, 
S.~Redfield$^{24}$,
and
M.~Tamura$^{4,13,15}$
\newauthor
\\
$^{1}$Sub-department of Astrophysics, Department of Physics, University of Oxford, Oxford, OX1 3RH, UK\\
$^{2}$Space Research Institute, Austrian Academy of Sciences, Schmiedlstrasse 6, A-8041 Graz, Austria\\
$^{3}$Dipartimento di Fisica, Universit\`{a} di Torino, via P. Giuria 1, 10125 Torino, Italy\\
$^{4}$Department of Astronomy, University of Tokyo, 7-3-1 Hongo, Bunkyo-ku, Tokyo 113-0033, Japan\\
$^{5}$Department of Earth and Space Sciences, Chalmers University of Technology, Onsala Space Observatory, 439 92 Onsala, Sweden\\
$^{6}$Leiden Observatory, University of Leiden, PO Box 9513, 2300 RA, Leiden, The Netherlands\\
$^{7}$Rheinisches Institut f\"{u}r Umweltforschung an der Universit\"{a}t zu K\"{o}ln, Aachener Strasse 209, 50931 K\"{o}ln\\
$^{8}$Instituto de Astrof\'{i}sica de Canarias, 38205 La Laguna, Tenerife, Spain\\
$^{9}$Departamento de Astrof\'{i}sica, Universidad de La Laguna, 38206 La Laguna, Tenerife, Spain\\
$^{10}$INAF - Osservatorio Astrofisico di Catania, Via S. Sofia 78, 95123 Catania, Italy\\
$^{11}$Astrophysics Group, Cavendish Laboratory, University of Cambridge, J. J. Thomson Avenue, Cambridge CB3 0HE, UK\\
$^{12}$Dipartimento di Fisica e Astronomia "Galileo Galilei", Universit\'a di Padova, Vicolo dell'Osservatorio 3, 35122, Padova, Italy\\
$^{13}$Astrobiology Center, 2-21-1 Osawa, Mitaka, Tokyo 181-8588, Japan \\
$^{14}$JST, PRESTO, 2-21-1 Osawa, Mitaka, Tokyo 181-8588, Japan \\
$^{15}$National Astronomical Observatory of Japan, 2-21-1 Osawa, Mitaka, Tokyo 181-8588, Japan \\
$^{16}$Stellar Astrophysics Centre, Department of Physics and Astronomy, Aarhus University, Ny Munkegade 120, DK-8000 Aarhus C, Denmark \\
$^{17}$INAF - Osservatorio Astronomico di Padova, Vicolo dell'Osservatorio 5, Padova, IT-3512, Italy\\
$^{18}$Institute of Planetary Research, German Aerospace Center, Rutherfordstrasse 2, 12489 Berlin, Germany\\
$^{19}$Department of Astronomy and McDonald Observatory, University of Texas at Austin, 2515 Speedway, Stop C1400, Austin, TX 78712, USA\\
$^{20}$ Department of Earth and Planetary Science, The University of Tokyo, 7-3-1Hongo, Bunkyo-ku, Tokyo 113-0033, Japan \\
$^{21}$Th\"{u}ringer Landessternwarte Tautenburg, Sternwarte 5, 07778 Tautenburg, Germany\\
$^{22}$Center for Astrophysics, Harvard \& Smithsonian, 60 Garden Street, Cambridge, MA 02138, USA\\
$^{23}$Space Telescope Science Institute, 3700 San Martin Drive, Baltimore, MD 21218, USA\\
$^{24}$Astronomy Department and Van Vleck Observatory, Wesleyan University, Middletown, CT 06459, USA 
}

\date{Last updated 2015 May 22; in original form 2013 September 5}

\pubyear{2015}

\begin{document}
\label{firstpage}
\pagerange{\pageref{firstpage}--\pageref{lastpage}}
\maketitle

\begin{abstract}
We present a detailed analysis of HARPS-N radial velocity observations of \target, a young and active star in the Praesepe cluster, which hosts a transiting planet with a period of 1.7~days. 
We model the activity-induced radial velocity variations of the host star with a multi-dimensional Gaussian Process framework and detect a planetary signal of \kb\, which matches the transit ephemeris, and translates to a planet mass of \mpb.
{  We perform a suite of validation tests to confirm that our detected signal is genuine.}
This is the first mass measurement for a transiting planet in a young open cluster.
The relatively low density of the planet, \denpb, implies that \target b retains a significant volatile envelope. 
{  We estimate that the planet is losing its atmosphere at a rate of $10^{11}-10^{12}\,{\rm g\,s^{-1}}$ due to the high level of radiation it receives from its host star.}
\end{abstract}



\section{Introduction}

Theoretical evolution models predict that the most significant changes in the bulk and orbital parameters of exoplanets occur in the first few hundred Myr of their evolution \citep[e.g.,][]{Adams2006,Kubyshkina2018b,Raymond2009}. 
Planets orbiting stars in young open clusters are thus particularly valuable tests of these models.
The exquisite photometry collected by the \ktwo\ space mission \citep{Howell2014} and its observing strategy focused on the Ecliptic plane have enabled the detection of the first transiting planet candidates in star forming regions and young stars \citep[e.g.,][]{David2016b,David2016a,David2019,Libralato2016,Mann2016b,Mann2016a,Mann2017,Mann2018,Pepper2017,Livingston2018a,Livingston2019}, but none so far has mass measurements. 
Recent studies show that these young transiting exoplanets seem to be larger than their counterparts with similar periods orbiting more evolved stars \citep[][]{Mann2016a}. This suggests that photoevaporation by the host star plays an important role in shaping the planet atmosphere in the first few Gyr \citep[as predicted by e.g.,][]{Owen2013}. However, expected evaporation rates depend strongly on planet mass, so measuring masses for these young transiting planets is important to test this scenario further.

This paper presents the first firm RV confirmation of a transiting planet in a young open cluster. \target\ (EPIC\,211990866, $\alpha_{\rm J2000}$= 08:38:24.30,
$\delta_{\rm J2000}$= +20:06:21.83) is a bright ($V=10.52$\,mag) G-dwarf member
\citep{Hillenbrand2007} of the Praesepe cluster (NGC\,2632, M44), which has an estimated age of $700$--$800$\,Myr and distance of $\sim 180$\,pc \citep{Brandt2015,Bossini2019,Salaris2004,vanLeeuwen2009}. The transits of \target b, were discovered independently by \citet{Pope2016} and \citet[][hereafter \citetalias{Mann2017}]{Mann2017} in \ktwo\ campaign 5 data, though only the latter identified the host star as a Praesepe member. 
Analysis of the \ktwo\ light curve alongside optical and infrared spectroscopy and adaptive optics imaging enabled \citetalias{Mann2017} to rule out most false positive scenarios and statistically validate the planetary nature of K2-100b, alongside 6 other Praesepe candidates orbiting fainter stars. With a period of 1.67\,d and $\sim 800$\,ppm transits, which implies a planet radius of $\sim 3.8\,R_\oplus$, K2-100b is a hot Neptune, and its bright host star made it a good candidate for further characterisation.

The RV follow-up of planets in young open clusters is challenging because their host stars rotate rapidly and are magnetically active. This gives rise to quasi-periodic variations in the apparent stellar RV, which can be very difficult to disentangle from the planetary signal(s). Gaussian Process Regression (GPR) can be used to model activity signals in RV data \citep[see e.g.][]{Haywood2014,Grunblatt2015}. This
approach is even more powerful when complementary activity indicators extracted from the spectra are modelled alongside the RVs, as in the framework developed by \citet[][hereafter \citetalias{Rajpaul2015}]{Rajpaul2015}. In this paper, we used the framework of \citetalias{Rajpaul2015} to analyse 
RV observations of K2-100 and detect the reflex motion of the star induced by the transiting planet at the $> 3\,\sigma$ level, despite the fact that the latter is of considerably lower amplitude than the activity-induced variations. 

\section{Observations}

\subsection{Photometry}
\label{sec:photometry}

\ktwo\ observed \target\ as part of its Campaign 5 (C5,  2015-04-27 UTC to 2015-07-10 UTC) in long-cadence mode (30 min). This star was re-observed by \ktwo\ in short cadence  (1\,min) mode on its Campaign 18 (C18, 2018-05-12 UTC to 2018-07-02 UTC).
We downloaded the K2SFF \citep{Vanderburg2014} light curve for C5 from the Mikulski Archive for Space Telescopes (\url{https://archive.stsci.edu/k2/}).
We used the \texttt{lightkurve} package \citep{lightkurve} to obtain the C18 \ktwo\ light curve.
We corrected for systematics using the pixel level decorrelation (PLD) as implemented in the \texttt{lightkurve} package.

\citet{Stefansson2018} performed a ground-based photometric follow-up of \target. They used the Engineered Diffuser instrument on the Astrophysical Research Council Telescope Imaging Camera (ARCTIC) 
imager located at the ARC 3.5 m Telescope at Apache Point Observatory. We downloaded the available public light curve from the online version of \citet[][]{Stefansson2018} to use it in the analysis presented in Sect. \ref{sec:modelling}.

{  We observed three transits} of \target\ with the MuSCAT2 multicolour photometer \citep{Narita2019} installed in the Carlos Sanchez Telescope (TCS) in the Teide observatory on the {nights of 2018-12-28 UTC, 2019-01-02 UTC, and 2019-01-22 UTC}. All observations covered from 2 to 3.2 hours around the expected mid-transit time, and were carried simultaneously in the $r'$, $i'$, and $z'$ passbands with a common exposure time of 10 seconds. The photometry was done with the MuSCAT2 pipeline based on \texttt{PyTransit} \citep{pytransit} and \texttt{LDTk} \citep{Parviainen2015b}.

We searched for transit timing variations (TTVs) using \texttt{PyTV} (Python Tool for Transit Variations, Korth 2019, {in prep.}).  
We detected no TTVs; therefore, our results are consistent with a constant period model. This result, together with the precise ephemeris, implies that \target\ can be efficiently scheduled for future follow-up observations.

\subsection{Spectroscopy}

We acquired 78 high-resolution ($R$$\approx$$115\,000$) spectra of \target\ with the HARPS-N spectrograph mounted at the 3.58-m Telescopio Nazionale Galileo at Roque de Los Muchachos observatory (La Palma, Spain), as part of the observing programs CAT15B\_35  (PI: Deeg), CAT15B\_79 (PI: Palle), and ITP16\_6 (PI: Malavolta). 
We processed the data using the dedicated HARPS-N pipeline and extracted the RVs by cross-correlating the HARPS-N spectra with a G2 numerical mask. We also extracted the Ca\,$\sc \mathrm{II}$ activity indicator $\log R'_\mathrm{HK}$ assuming a B-V\,=\,0.583.  Table~\ref{tab:rvsk2100} reports the HARPS-N RVs and their uncertainties along with the full-width at half maximum (FWHM) and the bisector inverse slope (BIS) of the cross-correlation function (CCF), $\log R_\mathrm{HK}$, and the signal-to-noise ratio (S/N) per pixel at 5500\,\AA. For the analysis presented in Sect.~\ref{sec:modelling}, we removed 5 RV data points with a relative low signal-to-noise (S/N\,<\,12). 

\section{Data analysis}

\subsection{Stellar parameters}
\label{sec:stellar}

We determined the spectroscopic parameters of \target\ from the co-added HARPS-N spectrum using the software Spectroscopy Made Easy \citep[\texttt{SME, version 5.22;}][]{Piskunov2017, Valenti1996} along with ATLAS12 model atmospheres \citep[][]{Kurucz2013} and atomic/molecular parameters from the VALD database \citep{Ryabchikova2015}. 
The effective temperature \teff, surface gravity \logg, iron abundance [Fe/H], and projected rotational velocity \vsini\ were measured following the same techniques described in, e.g., \citet{Fridlund2017}, \citet{Gandolfi2017}, and \citet{Persson2018}. 
The micro- (\vmic) and macro-turbulent (\vmac) velocities were fixed through the empirical calibration equations of \citet{Bruntt2010} and \citet{Doyle2014}.
As a sanity check, we also carried out an independent spectroscopic analysis using the package \texttt{specmatch-emp} \citep{Yee2017}. This code compares the observed spectrum with a library of $\sim$400 FGKM template spectra and minimises the differences between the observed and the library data. The derived spectroscopic parameters agree  within 1-sigma with those found by \texttt{SME}.

Following the method described in \citet{Gandolfi2008}, we measured the interstellar extinction along the line of sight to the star and found that it is consistent with zero. We derived the stellar mass, radius, and age using the on-line interface \texttt{PARAM-1.3} (\url{http://stev.oapd.inaf.it/cgi-bin/param}) and \texttt{PARSEC} stellar tracks and isochrones \citep{Bressan2012}, along with the visual magnitude \citep[V=10.56;][]{Mermilliod1987}, the \texttt{GAIA} parallax \citep[$\pi$=5.2645\,$\pm$\,0.0678 mas;][]{Brown2018}, and our effective temperature and iron abundance measurements. The derived stellar parameters are listed in Table~\ref{tab:parsk2100}.
We note that the inferred supersolar metallicity of \target\ ($[{\rm Fe/H}] = 0.22 \pm 0.09$) is consistent with previous values measured for Praesepe stars \citep[][]{Boesgaard2013,Pace2008}.

\subsection{Stellar density analysis}
\label{sec:stellardensity}

\begin{figure}
    \centering
    \includegraphics[width=0.47\textwidth]{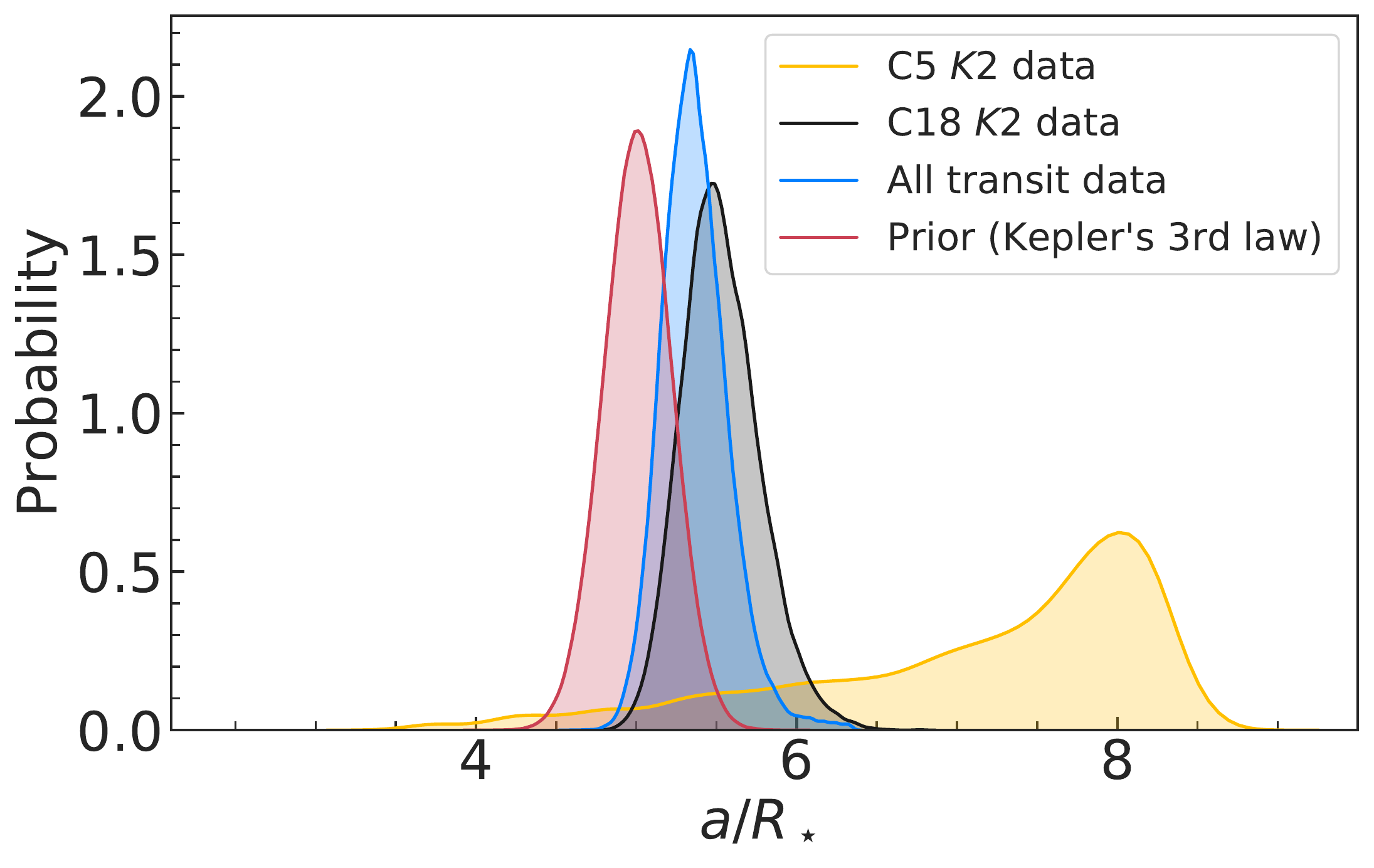}\\
    \caption{  Posterior distribution for $a/R_\star$ for different analyses.
    The posterior distribution for $a/R_\star$ fitting only \ktwo\ C5 data and \ktwo\ C18 data are shown in yellow and black, respectively.  Blue shows the posterior distribution for $a/R_\star$ fitting all available transits. We also show the prior on $a/R_\star$ using the derived stellar parameters in Sect.~\ref{sec:stellar} and the planetary orbital period in red.}
    \label{fig:bposteriors}
\end{figure}

\citetalias{Mann2017} and \citet{Livingston2018b} noticed that \target's stellar density coming from the light curve analysis ({  assuming a circular orbit}) differs from that from the spectroscopic parameters.
{  
This could be explained by a mischaracterised host star or an eccentric orbit.
 We discard the possibility that the star is mischaracterised given that our independent stellar parameter estimation is in agreement with the values reported by \citetalias{Mann2017} and \citet{Livingston2018b}. We also discard a significantly eccentric orbit given that the circularisation time of \target b's orbit \citep[$\approx$ 20 Myr, following][]{Jackson2008} is significantly smaller than the system age.

We found out that this discrepancy was caused by a wide posterior distribution for $a/R_\star$ when analysing \ktwo\ C5 data only. Figure~\ref{fig:bposteriors} shows the posterior distribution for the scaled semi-major axis, $a/R_\star$ \citep[that relates directly with stellar density, see e.g.,][]{Winn2010}, by fitting C5 \ktwo\ data only, C18 \ktwo\ data only, and also by fitting all available transits. We set uniform priors on $a/R_\star$ for all cases. When fitting the C5 \ktwo\ data, the MCMC converges to a solution which produces a wide posterior for $a/R_\star$ with median and 68\% credible interval given by $7.40^{+0.70}_{-1.75}$. This solution translates to a stellar density of $2.73^{+0.85}_{-1.62} \, \mathrm{g\,cm^{-3}}$. These values are similar to the values reported by \citetalias{Mann2017} and \citet{Livingston2018b}.
When fitting all available transits, the MCMC sampling converges to a narrower posterior distribution with a inferred value of $a/R_\star = 5.36^{+0.25}_{-0.20}$ (we note that this value is still inside the posterior distribution found by fitting only C5 \ktwo\ data). This value gives a stellar density of $\rho_\star = 1.04 \pm 0.15 \, \mathrm{g\,cm^{-3}}$ which is consistent with the value derived in Sect.~\ref{sec:stellar} (see Fig.~\ref{fig:bposteriors}).
We note that when fitting the C18 \ktwo\ alone we also get a $a/R_\star$ which is consistent with the expected value of $a/R_\star$ from Kepler's third law and the stellar parameters derived in Sect-~\ref{sec:stellar}.
The new analysis including all available transits suggests that the planetary orbit is nearly circular, therefore we assume a circular orbit for \target b's  in the rest of the manuscript.
In order to speed-up convergence for the final analysis presented in Sect.~\ref{sec:modelling}, we used the derived stellar parameters and Kepler's third law to set a Gaussian prior on $a/R_\star$ (see Fig.~\ref{fig:bposteriors}).

We note that the inferred $a/R_\star$ has a direct effect on the geometry on the system. For instance, the orbital inclination, planet radius, and other derived quantities differ from those reported in \citetalias{Mann2017} and \citet[][]{Livingston2018b}.

}

\begin{table}
\centering
  \caption{Stellar parameters. \label{tab:parsk2100}}  
  \begin{tabular}{lcc}
  \hline
  Parameter & Value & Source  \\
  \hline
    Stellar mass $M_{\star}$ ($M_\odot$)  &  \smass[] & This work \\
    Stellar radius $R_{\star}$ ($R_\odot$)  & \sradius[]  & This work \\
    \vsini (\kms)  & $14 \pm 2$ & This work  \\
    Stellar density $\rho_\star$ (g\,cm$^{-3}$)  &  \densspb[]  & This work \\   
    Effective Temperature $\mathrm{T_{eff}}$ (K)  & \stemp[] & This work \\
    Surface gravity $\log g_\star$ (cgs)  & $4.33 \pm 0.10 $  & This work\\ 
    Iron abundance [Fe/H] (dex)  & $0.22 \pm 0.09$ & This work \\
    Star age (Myr)  & $750^{+4}_{-7}$ & B19 \\
    Spectral type & G0V & PM13 \\
    \hline
  \end{tabular}
  \begin{tablenotes}\footnotesize
  \item \emph{Note:} B19 - \citet[][]{Bossini2019}, PM13 - \citet[][]{Pecaut2013}.
\end{tablenotes}
\end{table}

\subsection{Planet validation}

\target b was first validated by \citet{Mann2017}, who computed a false-positive probability (FPP) of 0.36\% using the {\tt vespa} software package \citep{Morton2012}. \citet{Livingston2018b} subsequently analysed the \ktwo\ data \citep[as processed by \href{https://exofop.ipac.caltech.edu/k2/edit_target.php?id=211990866}{\tt k2phot};][]{Petigura2015}
and obtained a slightly higher FPP of 1.2\%, just above their validation threshold of 1\%. This disagreement in FPP is comparatively small, and likely results from the use of different photometric pipelines, as well as stellar parameter estimates.
We have used the new information contained in the short cadence \ktwo\ C18 photometry of \target\ and our simultaneous multi-band MuSCAT2 photometry to revisit the FPP of \target b. The short cadence \ktwo\ data put tighter constraints on the transit shape than was possible with the long cadence data from C5, which in turn has a significant impact on the FPP. We now obtain an extremely low FPP of $\lesssim 10^{-6}$ for \target b using {\tt vespa}.

We can also independently constrain the possibility of various false positive scenarios by measuring $r_p \equiv R_p/R_\star$ in different bandpasses \citep[see e.g., ][]{Parviainen2019}. 
{  We performed a fit to all our available flattened transits allowing for a free $r_p$ for each band with uniform priors between [0,0.05]. We got $r_{p,K2} =$ \rrbkepler, $r_{p,{\rm ARCTIC}} =$ \rrbG, $r_{p,r} =$ \rrbbandr, $r_{p,i} =$ \rrbbandi, $r_{p,z} =$ \rrbbandz; the agreement of $r_p$ in these bandpasses is inconsistent with most false positive scenarios, thus confirming the {\tt vespa} result. }

\subsection{RV analysis using multi-dimensional GP}
\label{sec:gpapproach}

In this work we use the GP framework presented by  \citetalias{Rajpaul2015} to model  the RV data along with the \logr\ and BIS. Briefly, this approach assumes that all stellar activity signals can be modelled by the same latent variable $G(t)$ (and its derivatives) which is described by a zero-mean GP and a covariance function $\gamma$. Following \citetalias{Rajpaul2015}, the RV, \logr\ and BIS time-series can be modelled as
\begin{equation}
    \begin{matrix}
    \Delta RV & = & V_c G(t) + V_r \dot{G}(t), \\
    \log R'_{\rm HK} & = & L_c G(t), \\
    BIS & = & B_c G(t) + B_r \dot{G}(t),
\end{matrix}
\end{equation}
respectively. The variables $V_c$, $V_r$, $L_c$, $B_c$ and $B_r$ are free parameters which relate the individual time series to an underlying Gaussian Process $G(t)$. The GP itself is a latent (unobserved) variable, which can be loosely interpreted as representing the projected area of the visible stellar disc that is covered in spots or active regions at a given time. The GP is assumed to have zero mean and covariance matrix {\bf K}, where $K_{ij}=\gamma(t_i,t_j)$. Following \citetalias[][]{Rajpaul2015}, we  adopt the quasi-periodic covariance function
\begin{equation}
    \gamma(t_i,t_j) = \exp 
    \left[
    - \frac{\sin^2[\pi(t_i - t_j)/P_{\rm GP}]}{2 \lambda_{\rm P}^2}
    - \frac{(t_i - t_j)^2}{2\lambda_{\rm e}^2}
    \right],
    \label{eq:gamma}
\end{equation}
where $P_{\rm GP}$ is the period of the activity signal, $\lambda_p$ the inverse of the harmonic complexity, and $\lambda_e$ is the long term evolution timescale. This choice of covariance function is widely used to model stellar activity signals in both photometry and RVs (see e.g. \citealt{Aigrain2012,Haywood2014} and \citetalias{Rajpaul2015}). The full expressions for the covariance between the three types of observations are given in \citetalias{Rajpaul2015}.

\subsection{RV and transit modelling}
\label{sec:modelling}

We used the open source code \href{https://github.com/oscaribv/pyaneti}{\pyan} \citep[][]{pyaneti} to model the light curve and RV data. We modified \pyan's public version to allow for multi-band transit and GP analyses. 
We also implemented the multi-dimensional GP approach described in Sect. \ref{sec:gpapproach} and \citetalias{Rajpaul2015}. 
 
We used  \href{https://github.com/oscaribv/exotrending}{\texttt{exotrending}} \citep{exotrending} to isolate each transit and to remove long term trends in the light curves as described in \citet{Barragan2018a,Barragan2018b}. We re-sampled the model over ten steps to account for the long-cadence (30min, C5) \ktwo\ data \citep{Kipping2010}. 
We did not re-sample the model for \ktwo\ and ground-based short-cadence data.
We assumed that the difference of transit depth between different bands is negligible; therefore, we fit for a single radius ratio $R_p/R_\star$ for all the bands. We fitted for the limb darkening parameters for each band using uniform priors and the parametrisation described by \citet[][]{Kipping2013}.
We have assumed a circular orbit (See Sect.~\ref{sec:stellardensity}).

We performed a joint fit of all transits together with the RV, \logr, and BIS time-series using the approach presented in Sect.~\ref{sec:gpapproach}.
A summary of the fitted parameters and priors are presented in Table~\ref{tab:parsk2100b}. We used 500 chains to sample the parameter space (38 free parameters). For the burning-in phase we used the last 5000 of converged chains with a thin factor of 10, leading to a final number of 250,000 independent points for each fitted parameter.

\section{Results and Discussion}
\label{sec:results}

\begin{figure*}
    \centering
    \includegraphics[width=1\textwidth]{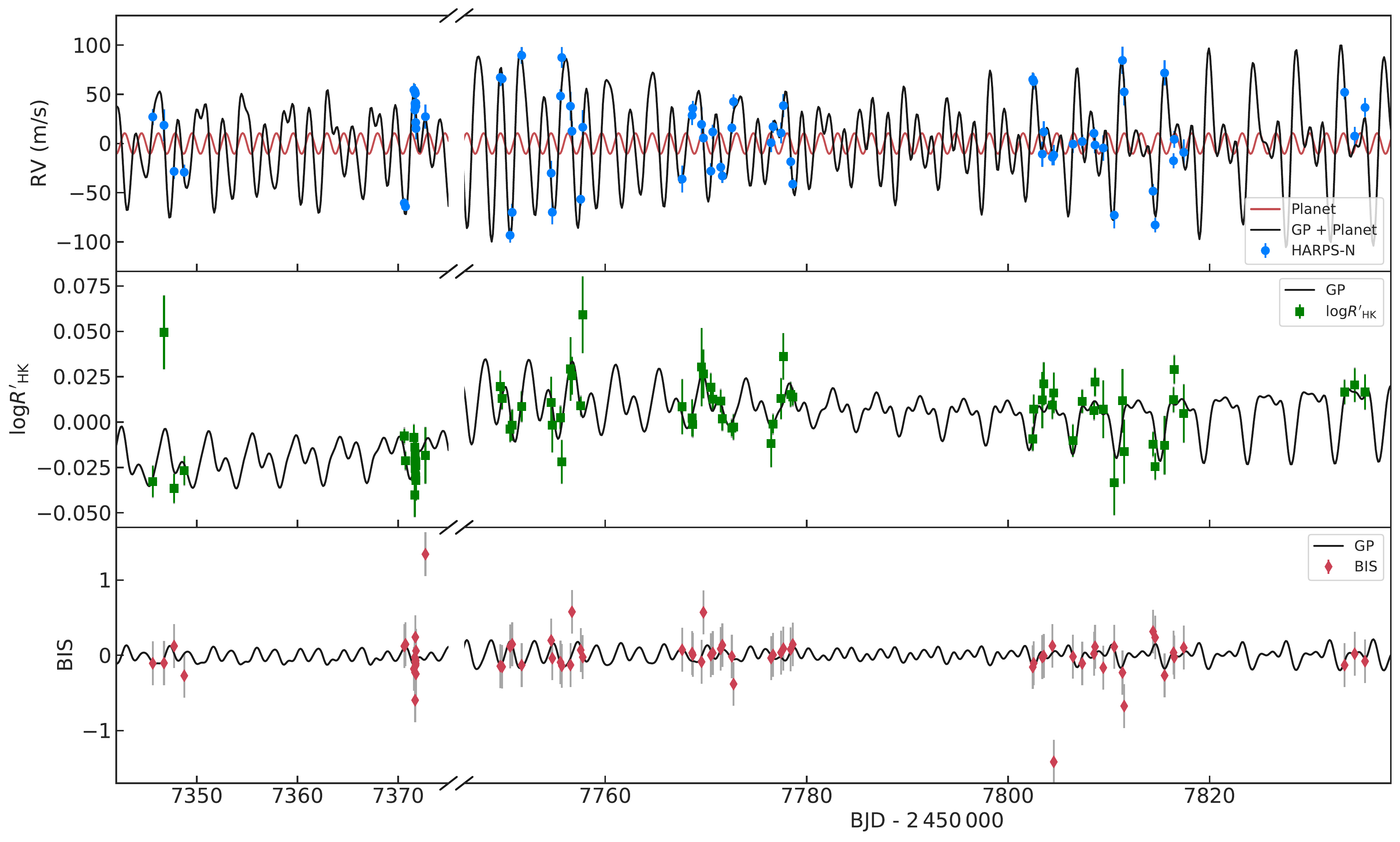}\\
    \caption{Radial velocity (top), \logr\ (middle) and BIS (bottom) time-series. All time-series have been corrected by the inferred offset. Inferred models are presented as solid continuous lines. Measurements are shown with filled symbols with error bars.  Grey error bars account for the jitter. We note that there is a gap between 7375 and 7746 BJD - 2\,450\,000 where there were no measurements.}
    \label{fig:timeseries}
\end{figure*}

\begin{figure}
    \centering
    \includegraphics[width=0.47\textwidth]{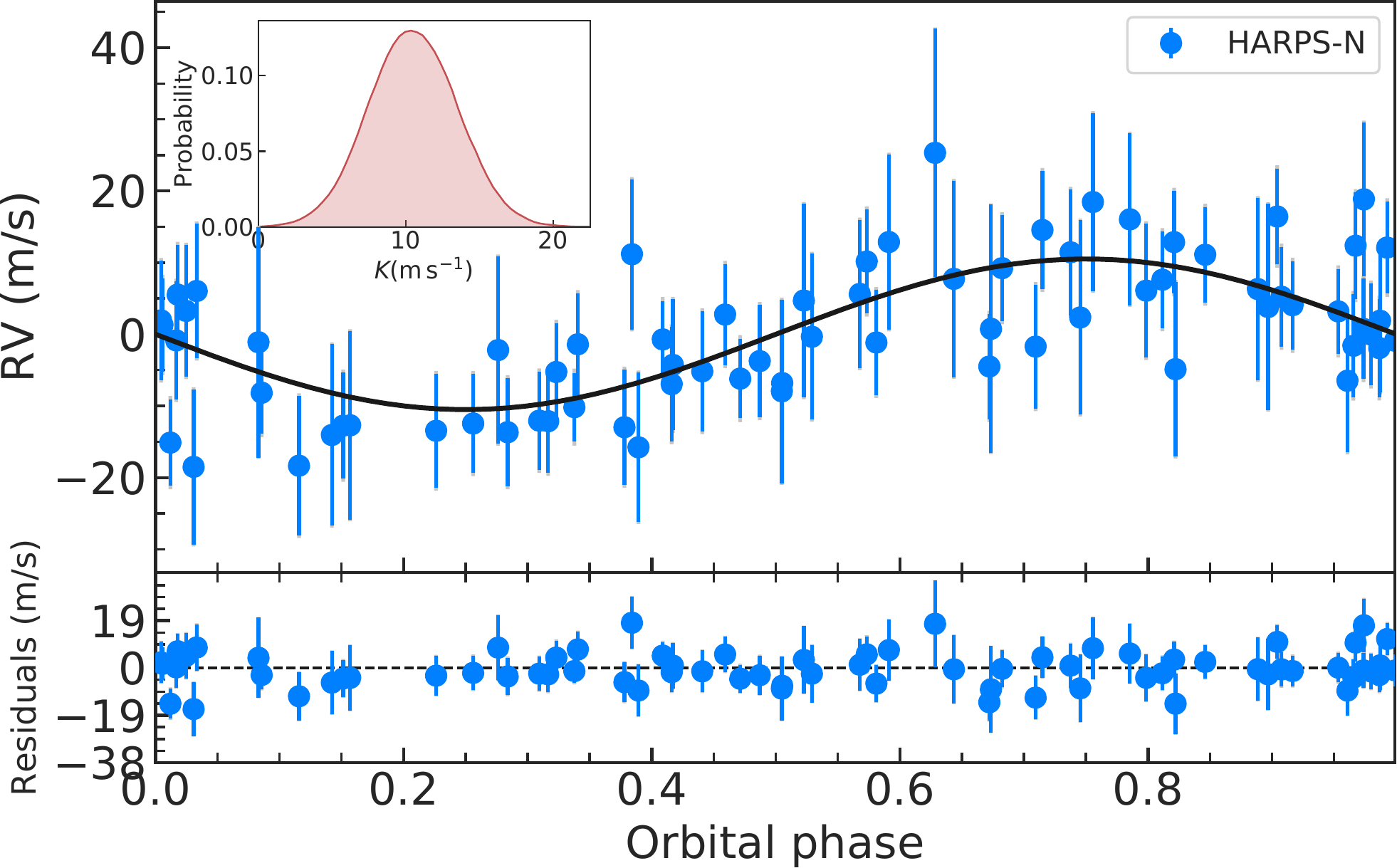}\\
    \caption{RV curve of \target\ folded to the orbital period of \target b. HARPS-N data (blue circles) are shown following the subtraction of the instrumental offset and GP model.  Grey error bars account for the jitter. The Keplerian solution is shown as a solid line. Top-left inset displays the posterior distribution for $K$.
    }
    \label{fig:rvplanet}
\end{figure}

\begin{figure}
    \centering
    \includegraphics[width=0.47\textwidth]{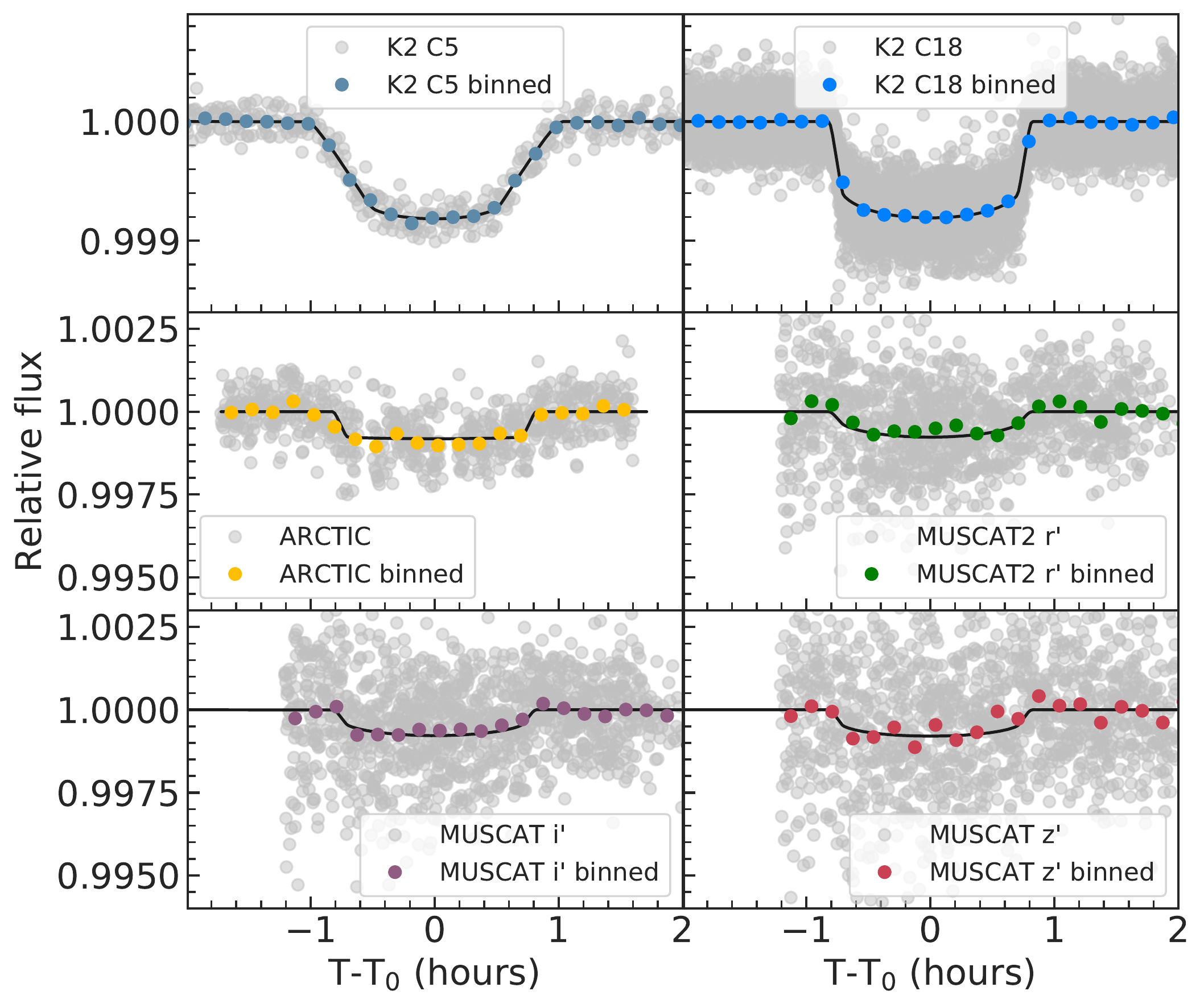}\\
    \caption{\target b transits. Each panel shows a flattened light curve from different instruments  folded to the orbital period of \target b. Black lines show the best-fitting transit models.}
    \label{fig:transits}
\end{figure}

 Figure~\ref{fig:timeseries} shows the RV, \logr\, and BIS time-series together with the inferred models. We also show the phase-folded RV and transit models, along with the data points in Figures ~\ref{fig:rvplanet} and \ref{fig:transits}, respectively.
 We inferred a planetary induced RV semi-amplitude of \kb, which translates into a planet mass of \mpb. Other parameter estimates are presented in Table~\ref{tab:parsk2100b}.

We note that we also analysed the RV data set with standard RV analysis techniques, such as Fourier decomposition \citep[e.g.,][]{Barragan2018a,Pepe2013} and GPs trained with photometry \citep[e.g.,][]{Barragan2018b,Malavolta2018}. We found hints of the induced Doppler signal with a significance $\lesssim$ 2-sigma.
This shows that the simultaneous regression of the activity/asymmetry indicators play a fundamental role to measure the Doppler semi-amplitude with higher precision.

As a first check of the validity of our detection, we compare the Bayesian Information Criteria \citep[BIC; see e.g.,][]{Burnham2002}.
We repeated the analysis presented in Sect.~\ref{sec:modelling} by fitting a model with and without planet. We model only the RV-related time-series, i.e., with no transit modelling.
For the fit with planet, we set priors on the ephemeris coming from the transit analysis.
We conclude that the model including the planet signal is strongly preferred over the model without it with a $\Delta {\rm BIC} = 26$.

\citet{Rajpaul2016} showed that spurious RV detection of planets around active stars can  arise due to a combination of complex activity models and the window function of the observations. To check that this is not the case here, we created 250 synthetic RV,  \logr, and BIS time-series using the best-fit GP model, with no planet in the RV data set. 
We added white noise to each point from a Gaussian distribution with standard deviation as the nominal error bar of each data point. 
We ran an MCMC fit as the one described in Sect. \ref{sec:modelling} (without transit data) for each data set, allowing for an RV signal with priors on the ephemeris of the planet. These simulations give rise to a "detection" (we define a "detection" as a signal with a significance $> 2$-sigma) only in 0.4\% of the cases.
We then repeat the experiment creating 250 more mock data sets, but this time injecting a coherent signal with an amplitude of 10\ms in the RV data set and same ephemeris as \target b. For this case we have a "detection" on $90$\% of the runs.
These results suggest that the planetary signal we detected in the real data is genuine. 

As a further test of the reliability of our detection, we also extracted the RV measurements with a K5 numerical mask, and repeated the analysis presented in Sect. \ref{sec:modelling}. We found an amplitude of $K = 12.4 \pm 3.5$ \ms\ which is within $1\,\sigma$ of the value obtained with the RVs extracted using the fiducial G2 mask.

\begin{table*}
  \caption{\target b parameters. \label{tab:parsk2100b}}  
  \begin{tabular}{lcc}
  \hline
  Parameter & Prior$^{(\mathrm{a})}$ & Value$^{(\mathrm{b})}$  \\
  \hline
  \multicolumn{3}{l}{\emph{ \bf Model Parameters for  \target b}} \\
    \noalign{\smallskip}
    Orbital period $P_{\mathrm{orb}}$ (days)  &  $\mathcal{U}[1.6737 , 1.6740]$ & \Pb[] \\
    Transit epoch $T_0$ (BJD - 2,450,000)  & $\mathcal{U}[  7140.70 , 7140.75 ]$ & \Tzerob[]  \\
    $e$  &  $\mathcal{F}[0]$ & 0  \\
    $\omega_\star $  &  $\mathcal{F}[\pi/2]$ & $\pi/2$  \\
    Scaled semi-major axis $a/R_{\star}$ &   $\mathcal{N} [ 5.01 , 0.21 ]$ & \arb[]\\
    Scaled planetary radius $R_\mathrm{p}/R_{\star}$ &  $\mathcal{U}[0,0.05]$ & \rrbKtwo[]  \\
    Impact parameter, $b$ &  $\mathcal{U}[0,1]$  & \bb[] \\
    Radial velocity semi-amplitude variation $K$ (m s$^{-1}$) &  $\mathcal{U}[0,50]$ & \kb[] \\
    GP Period $P_{\rm GP}$ (days) &  $\mathcal{U}[4,5.1]$ & \jPGP[] \\
    $\lambda_{\rm P}$ &  $\mathcal{U}[0.1,2]$ &  \jlp[] \\
    $\lambda_{\rm e}$ &  $\mathcal{U}[1,300]$ &  \jle[] \\
    $V_{c}$ (\kms)  &  $\mathcal{U}[0,0.1]$ & \jArvc \\
    $V_{r}$ (\kms) &  $\mathcal{U}[-1,1]$ & \jArvr \\
    $L_{c}$  &  $\mathcal{U}[0,1]$ & \jAhk \\
    $B_{c}$ (\kms) &  $\mathcal{U}[-1.5,1.5]$ & \jAbisc \\
    $B_{r}$ (\kms) &  $\mathcal{U}[-0.5,0.5]$ & \jAbisr \\
    Offset HARPS-N (\kms) & $\mathcal{U}[ 34.1998 , 34.5825]$ & \HARPSN[] \\
    Offset \logr\ & $\mathcal{U}[-4.5878 , -4.2885 ]$ & \RHK[]  \\
    Offset BIS (\kms) & $\mathcal{U}[-1.5568 , 0.6372 ]$ & \BIS[]  \\
    Jitter term $\sigma_{\rm HARPS-N}$ (\ms) & $\mathcal{U}[0,100]$ & \jHARPSN[] \\
    Jitter term $\sigma_{\log R'_{\rm HK}}$ & $\mathcal{U}[0,1]$ & \jRHK[] \\
    Jitter term BIS (\ms) & $\mathcal{U}[0,1000]$ & \jBIS[] \\
    Limb darkening $q_1$ for \ktwo\ C5 & $\mathcal{U}[0,1]$ & \qoneKtwo \\
    Limb darkening $q_2$ for \ktwo\ C5 & $\mathcal{U}[0,1]$ & \qtwoKtwo \\
        Limb darkening $q_1$ for \ktwo\ C18 & $\mathcal{U}[0,1]$ & \qoneKtwo \\
    Limb darkening $q_2$ for \ktwo\ C18 & $\mathcal{U}[0,1]$ & \qtwoKtwo \\
        Limb darkening $q_1$ for ARCTIC & $\mathcal{U}[0,1]$ & \qoneG\\
    Limb darkening $q_2$ for ARCTIC & $\mathcal{U}[0,1]$ & \qtwoG \\
        Limb darkening $q_1$ for MUSCAT2 r' & $\mathcal{U}[0,1]$ & \qonebandone \\
    Limb darkening $q_2$ for for MUSCAT2 r' & $\mathcal{U}[0,1]$ & \qtwobandone \\
        Limb darkening $q_1$ for MUSCAT2 i' & $\mathcal{U}[0,1]$ & \qonebandtwo \\
    Limb darkening $q_2$ for MUSCAT2 i' & $\mathcal{U}[0,1]$ & \qtwobandtwo \\
        Limb darkening $q_1$ for MUSCAT2 z' & $\mathcal{U}[0,1]$ & \qonebandone \\
    Limb darkening $q_2$ for MUSCAT2 z' & $\mathcal{U}[0,1]$ & \qtwobandtwo \\
     Jitter term $\sigma_{K2 {\rm C5}}$ ($\times 10^{-6}$) & $\mathcal{U}[0,1 \times10^{3}]$ & \jtrKtwo \\
    Jitter term $\sigma_{K2 {\rm C18}}$ ($\times 10^{-6}$) & $\mathcal{U}[0,1 \times10^{3}]$ & \jtrSC \\
    Jitter term $\sigma_{{\rm ARCTIC}}$ ($\times 10^{-6}$) & $\mathcal{U}[0,1 \times10^{3}]$ & \jtrG \\
    Jitter term $\sigma_{{\rm MUSCAT2} r'}$ ($\times 10^{-6}$) & $\mathcal{U}[0,1 \times10^{5}]$ & \jtrbandone \\
    Jitter term $\sigma_{{\rm MUSCAT2} i'}$ ($\times 10^{-6}$) & $\mathcal{U}[0,1 \times10^{5}]$ & \jtrbandtwo \\
    Jitter term $\sigma_{{\rm MUSCAT2} z'}$ ($\times 10^{-6}$) & $\mathcal{U}[0,1 \times10^{5}]$ & \jtrbandthree \\
    \hline
    \multicolumn{3}{l}{\emph{Derived parameters}} \\
  \noalign{\smallskip}
    Planet mass ($M_{\oplus}$)  & $\cdots$ & \mpb[] \\
    Planet radius ($R_{\oplus}$)  & $\cdots$ & \rpbKtwo[] \\
    Planet density (${\rm g\,cm^{-3}}$) & $\cdots$ & \denpb[] \\
    semi-major axis $a$ (AU)  & $\cdots$ & \ab[] \\
    Orbital inclination $i$ (deg)  & $\cdots$ & \ib[] \\
     Equilibrium temperature$^{(\mathrm{c})}$ $T_{\rm eq}$ ($K$)   & $\cdots$ & \Teqb[]  \\
    Insolation $F_{\rm p}$ ($F_{\oplus}$)   & $\cdots$ & \insolationb[] \\
     Planet surface gravity$^{(\mathrm{d})}$ (cm\,s$^{-2}$) & $\cdots$ & \grapb[] \\
    Planet surface gravity (cm\,s$^{-2}$)  & $\cdots$ & \grapparsb[] \\
    \hline
   \noalign{\smallskip}
  \end{tabular}
  \begin{tablenotes}\footnotesize
  \item \emph{Note} -- $^{(\mathrm{a})}$ $\mathcal{U}[a,b]$ refers to uniform priors between $a$ and $b$, $\mathcal{N}[a,b]$ to Gaussian priors with median $a$ and standard deviation $b$, and $\mathcal{F}[a]$ to a fixed value $a$.  
  $^{(\mathrm{b})}$  Inferred parameters and errors are defined as the median and 68.3\% credible interval of the posterior distribution.
 {  $^{(\mathrm{c})}$ Assuming albedo = 0.
 $^{(\mathrm{d})}$ Calculated from the scaled-parameters as in \citet[][]{Sotuhworth2007}.}
\end{tablenotes}
\end{table*}

\begin{figure}
\centering
\includegraphics[width=0.47\textwidth]{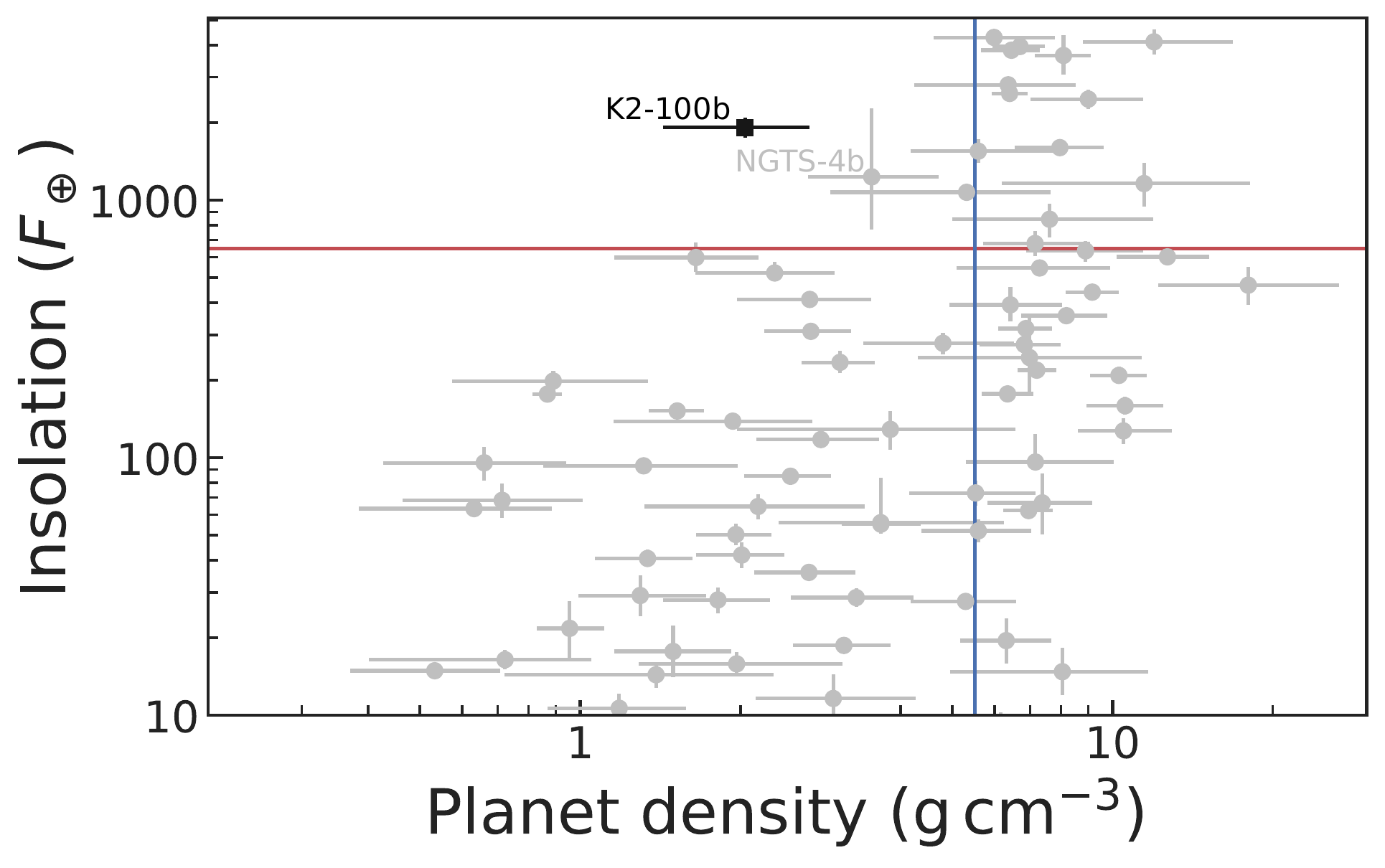}
\caption{  Planet density \emph{vs} insolation for small ($R_{\rm p} < 4\,R_\oplus$) transiting planets (grey circles). The location of \target b is marked with a black square. We also label NGTS-4b. Horizontal red line shows the insolation limit of $650 F_\oplus$ given by \citet[][]{Lundkvist2016}. Vertical blue line corresponds to Earth's density.
}
\label{fig:flux_density}
\end{figure}

{  Figure~\ref{fig:flux_density} shows a planet density \emph{vs} insolation plot for small planets ($R_{\rm p}< 4\, R_\oplus$) with masses measured to better than 50\% as listed in the TEPCAT catalogue \citep[][\url{http://www.astro.keele.ac.uk/jkt/tepcat/}]{tepcat}.
The plot also shows the limit of $650\,F_\oplus$ given by \citet[][]{Lundkvist2016} likely related to the presence/lack of a hydrogen-dominated atmosphere as a consequence of strong atmospheric escape. We find that for weakly irradiated planets ($< 650\,F_\oplus$), low (sub-Earth) densities are common, in contrast to highly irradiated for which most of the planets have densities equal or larger than that of the Earth, with only two exceptions: NGTS-4b \citep{west2019} and \target b. We discuss in more detail these two planets below. }

\begin{figure}
    \centering
    \includegraphics[width=0.45\textwidth]{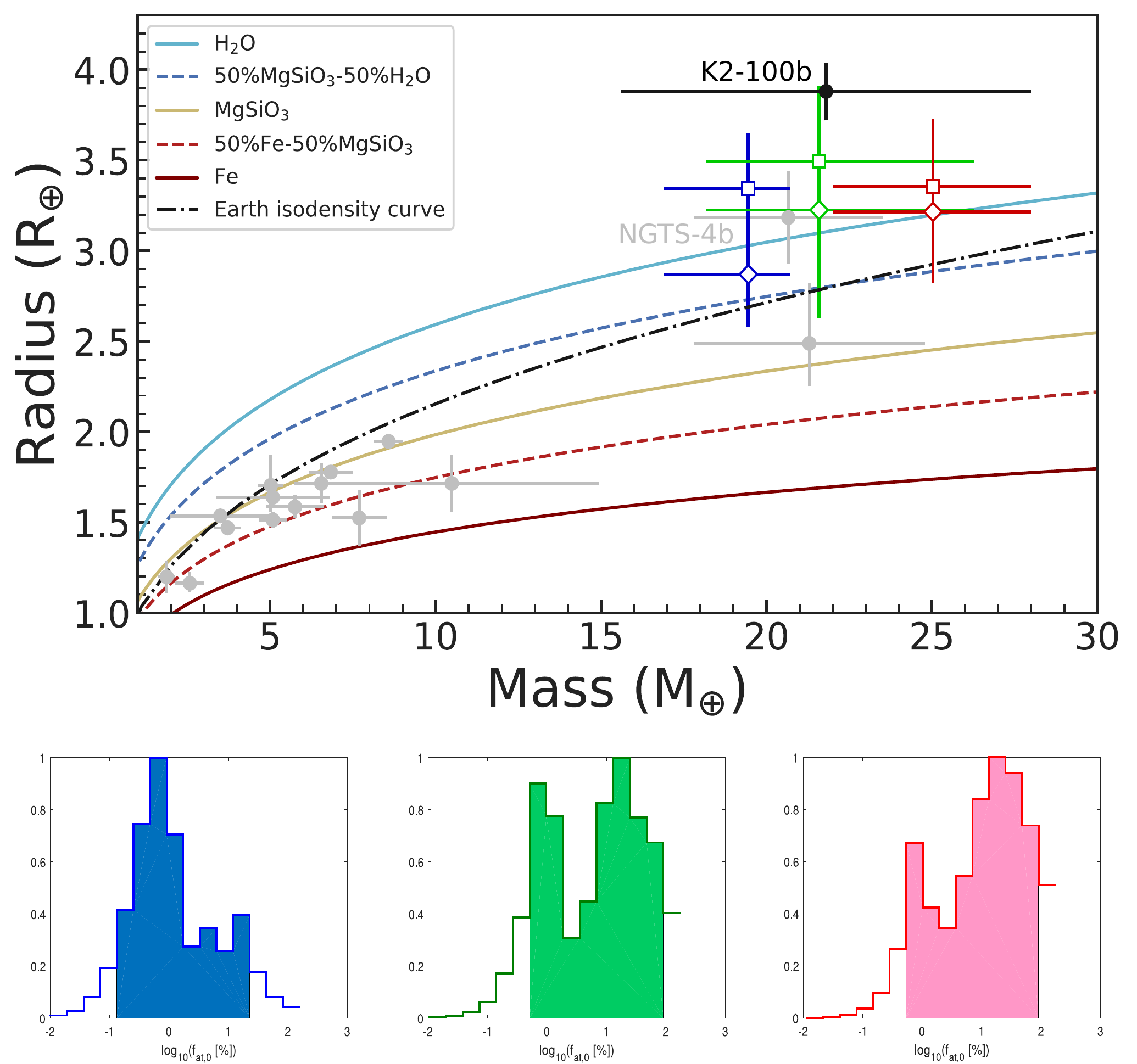}
    \caption{Top: Mass \emph{vs} radius diagram for small ($R_{\rm p} < 4\,R_\oplus$) planets which receive an insolation $> 650$ larger than the Earth (grey circles). The location of \target b is marked with a black circle. Its predicted planetary mass and radius at 2 and 5 Gyr is {  shown with empty squares and diamons}, respectively, with colours corresponding to different initial rotation rates XUV fluxes for the star: fast/high (red), moderate (green) and slow/low (blue) (see text for details). \citet[][]{Zeng2016}'s composition models are displayed with different colour lines. Bottom: posterior distributions obtained for the initial atmopsheric mass fraction $f_{\rm at,0}$ assuming the three different regimes of evolution of the stellar XUV flux. 
    The shaded areas correspond to the 68\% region of the credible interval of the posterior distribution.
    }
    \label{fig:mass_radius}
\end{figure}

Figure~\ref{fig:mass_radius} shows the position of \target b in a mass-radius diagram together with two-layer composition models by \citet{Zeng2016}.
With a mass of \mpb, a radius of \rpbKtwo, and a density of \denpb, we expect that \target b is a planet with a solid core with a significant volatile envelope.
{ 
Figure~\ref{fig:mass_radius} also shows all highly irradiated small planets from Figure~\ref{fig:flux_density}.
We find that all relatively low mass ($ \lesssim 10 M_\oplus$) planets have densities higher than that of the Earth and they are consistent with a composition made of different mixtures of iron and silicates. This can be explained by the fact that close-in, low-mass planets beyond this insolation limit are expected to lose their primordial H/He atmospheres \citep[e.g.,][]{Lundkvist2016}. 
For planets with higher masses ($ > 10 M_\oplus$), instead, bulk densities are typically lower than that of the Earth and compositions range from mixes of silicates and water to solid cores with volatile envelopes.
In fact, \citet[][]{west2019} argue that NGTS-4b's relative low density may be caused by a relatively high core mass, which enables the planet to retain a significant volatile envelope.}

Given the system's youth and short orbital separation, we model the past and future planetary atmospheric evolution, in particular to estimate if (and when) the planet will lose its envelope. To this end, we employed the planetary atmospheric evolution scheme described by \citet[][hereafter \citetalias{kubyshkina2018} and \citetalias{kubyshkina2019}, respectively]{kubyshkina2018,kubyshkina2019}. This is based on a combination of model grids and analytical approximations. They comprise models providing atmospheric mass-loss rates as a function of system parameters \citepalias{kubyshkina2018}, models enabling to estimate the atmospheric mass fraction as a function of planetary parameters \citep[i.e., radius, mass, equilibrium temperature;][]{johnstone2015a}, and the Mesa/MIST grid of stellar evolutionary tracks to account for the evolution of the stellar bolometric luminosity \citep{choi2016}. We model the past and future evolution of the stellar rotation period using a prescription similar to the empirical period-colour-age relation of \citet{mamajek2008}, modified to match the present-day rotation period, but with a free parameter $x$ allowing us to vary the spin-down rate prior to 2\,Gyr \citepalias[see][for details]{kubyshkina2019}. The instantaneous high-energy X-ray+EUV (XUV) emission of the host star is estimated from the rotation period following \citet{wright2011}, allowing us to explore a wide range of scenarios for the integrated XUV budget of the planet over its lifetime.

As described in \citetalias{kubyshkina2019}, we apply a Monte Carlo approach to fit the observed planetary radius, using the other system parameters and their {  uncertainties as inputs, finally obtaining probability distribution functions for $x$ and the initial atmospheric mass fraction (i.e., ratio between atmospheric mass and planetary mass at an age of 5\,Myr; $f_{\rm at,0}$) as output. Altogether, the input parameters of the Monte Carlo simulation are planetary mass, orbital separation, age of the system, stellar mass, and present-day rotation period\footnote{The planetary equilibrium temperature, which is one of the input parameters for extracting the mass-loss rates (by setting the lower boundary of the hydrodynamic modelling) and atmospheric mass fractions from the grids, is set by the orbital separation and stellar parameters, where the latter are derived from the MESA evolutionary tracks and the stellar mass.}. We then use the results to evolve the planetary atmosphere beyond its current age and up to 5\,Gyr, computing the planetary radius and $f_{\rm at}$ as a function of age and for three different ranges of $x$ corresponding to rotation rates at an age of 150\,Myr of less than 0.5\,days, between 0.5 and 3\,days, and more than 3 days. At an age of 150\,Myr, these rotation rates translate to XUV fluxes in the range 376--600, 117--376, and 13--117 times larger than the current solar XUV emission, respectively. Throughout, we assume a core density equal to Earth's bulk density, which sets the core radius.}

{  Figure~\ref{fig:mass_radius} shows the current position of the planet in the mass-radius diagram and those predicted to be possible at 2 and 5\,Gyr, for the three different ranges of $x$ we considered. Our results indicate that after 5\,Gyrs the planet is likely to lose a significant amount of its primordial hydrogen-dominated atmosphere, finally retaining between about 0.1 and 0.7\% of its mass in the atmosphere, depending on the evolutionary path of the stellar XUV emission and on planetary mass. In particular, for a planetary mass below about 20\,$M_{\oplus}$ it is unlikely that the planet will retain more than 0.1\% of its mass in the atmosphere and therefore its predicted radius at 5\,Gyr is close to the assumed core radius. In some cases, when considering planetary masses below $\sim$18\,$M_{\oplus}$, we reach the (almost) complete escape of the primary atmosphere before 2\,Gyr.}

In case the actual planetary mass is above about 20\,$M_\oplus$, the planet could still keep up to 0.7\% of its mass in the atmosphere, as shown in Fig.~\ref{fig:mass_radius}. 
This plot shows that if \target\ evolves as described by our fast rotator model, \target b should have a relatively high core mass, which is able to retain a significant volatile envelope. This could be similar to the case of NGTS-4b. On the other extreme, if \target\ evolves as a slow rotator, which is possible if the planet has a mass closer to the lower mass limit given by the RV measurements, \target b would end up as a core with an Earth-like density, similar to the other highly irradiated planets.

As shown by \citetalias{kubyshkina2019}, and illustrated on Figure~\ref{fig:mass_radius}, for a given stellar evolution scenario, the observed present-day radius of the planet can only be matched for a certain range of masses, which is within the mass range allowed (at the 1-sigma level) by our RV results. We were unable to fit the observed present-day radius with any atmospheric evolution scenario for planet masses below $\sim$15\,$M_{\oplus}$: at such low planetary masses, the atmosphere essentially escapes entirely before the age of Praesepe, even if we assume that the initial stellar XUV flux was rather low. 

Figure~\ref{fig:mass_radius} also presents the posterior distributions we obtained for the initial planetary atmospheric mass fraction $f_{\rm at,0}$ for the three different ranges of $x$ we considered. Larger XUV fluxes imply that more atmosphere has already escaped, so that the initial atmospheric mass fraction must have been larger (though the range of allowed values is also larger). {  Our results indicate that the planet may be subject to substantial atmospheric escape throughout most of its lifetime with the strongest escape happening during the first few hundred Myrs. In particular, for masses larger than about 20\,$M_\oplus$ atmospheric escape remains significant for Gyrs, implying that the planetary radius will keep decreasing, hence evolving, also after the first few hundred Myrs during which the planetary radius can decrease dramatically.
}

In all of the models that fit the available data, the planet is currently hosting an escaping atmosphere: {  using the code described by \citetalias{kubyshkina2018}}, we computed a series of hydrodynamic models of the planetary upper atmosphere for the range of planet parameters spanned by the evolution models that fit the observational constraints. These yield present-day atmospheric mass-loss rates in the range $10^{11}$--$10^{12}$\,g\,s$^{-1}$.

\section{Conclusions}

We showed how, by combining RV with activity indicators, we can disentangle planetary and activity RV variations for young active stars.
These results encourage the RV follow-up of young or active stars to be discovered with missions such as \emph{TESS} and \emph{PLATO}.

We measured a mass of \mpb\ for \target b, a \rpbKtwo\ planet transiting a star in the Praesepe cluster. 
We estimated that the relative high irradiation received by the planet implies that its atmosphere is currently evaporating. This makes \target\ an excellent laboratory to test photo-evaporation models.

\section*{Acknowledgments}

{\small
This paper includes data collected by the K2 mission. Funding for the K2 mission is provided by the NASA Science Mission directorate.
Based on observations made with the Italian Telescopio Nazionale Galileo (TNG) operated on the island of La Palma by the Fundaci\'on Galileo Galilei of the INAF (Istituto Nazionale di Astrofisica) at the Spanish Observatorio del Roque de los Muchachos of the Instituto de Astrofisica de Canaria.
Data for this paper has been obtained under the International Time Programme of the CCI (International Scientific Committee of the Observatorios de Canarias of the IAC).
This article is based on observations made with the MuSCAT2 instrument, developed by ABC, at Telescopio Carlos S\'anchez operated on the island of Tenerife by the IAC in the Spanish Observatorio del Teide.
These results are based on observations obtained with the
Apache Point Observatory 3.5-meter telescope, which is owned and operated by the Astrophysical Research Consortium.
O.B. and S.Ai. acknowledge support from the UK Science and Technology Facilities Council (STFC) under grants ST/S000488/1 and ST/R004846/1. 
J.K., S.G.  and A.P.H acknowledges support by Deutsche Forschungsgemeinschaft (DFG) grants PA525/18-1 and PA525/19-1 and HPA 3279/12-1  within the DFG Schwerpunkt SPP 1992, Exploring the Diversity of Extra-solar Planets. 
L.M. acknowledges support from  PLATO ASI-INAF agreement n.2015-019-R.1-2018.
S.Al. acknowledges the support from the Danish Council for Independent Research through the DFF Sapere Aude Starting Grant No. 4181-00487B, and the Stellar Astrophysics Centre which funding is provided by The Danish National Research Foundation (Grant agreement no.: DNRF106).
This work is partly supported by JSPS KAKENHI Grant Numbers JP18H01265, JP18H05439,  15H02063, and 18H05442 and JST PRESTO Grant Number JPMJPR1775.
M.C.V.F. and C.M.P. gratefully acknowledge the support of the  Swedish National Space Agency (DNR 174/18).
}

\bibliographystyle{mnras}
\bibliography{bibs} 

\appendix

\section{HARPS-N measurements}

\begin{table*}
\centering
  \caption{Radial velocity, activity and symmetry indicators measurements for \target. \label{tab:rvsk2100}}  
  \begin{tabular}{cccccccc}
  \hline
  Time & RV & $\sigma_{\rm RV}$ & CCF BIS & CCF FWHM & \logr & $\sigma_{\log { R'_{\rm HK}}}$ & S/N  \\
  (BJD$_\mathrm{TDB}$-2,450,000) & (\kms) & (\kms) & (\kms) & (\kms) & & \\
  \hline
7345.63288	&	34.4200	&	0.0080	&	21.1143	&	-0.1473	&	-4.4804	&	0.0086	&	51.3	\\
7346.74932	&	34.4116	&	0.0161	&	20.8156	&	-0.1438	&	-4.3982	&	0.0203	&	28.1	\\
7347.75251	&	34.3646	&	0.0074	&	20.8381	&	0.0827	&	-4.4842	&	0.0080	&	54.2	\\
7348.75941	&	34.3638	&	0.0076	&	20.9736	&	-0.3119	&	-4.4744	&	0.0078	&	54.2	\\
7351.70898	&	34.0950	&	0.0453	&	21.1253	&	0.6314	&	-4.3827	&	0.0741	&	11.3	\\
7351.73144	&	33.6321	&	0.1215	&	24.0350	&	1.0506	&	-3.9538	&	0.1979	&	2.3	\\
7352.72840	&	34.1035	&	0.0536	&	21.8887	&	0.4047	&	-4.4255	&	0.1046	&	9.1	\\
7352.74247	&	33.7331	&	0.0485	&	23.0619	&	2.5142	&	-4.2920	&	0.0653	&	10.4	\\
7370.61016	&	34.3325	&	0.0048	&	20.7341	&	0.0814	&	-4.4553	&	0.0039	&	82.2	\\
7370.72940	&	34.3289	&	0.0054	&	20.7907	&	0.1057	&	-4.4689	&	0.0048	&	70.9	\\
7371.55846	&	34.4474	&	0.0067	&	20.7710	&	-0.2209	&	-4.4561	&	0.0069	&	60.1	\\
7371.65315	&	34.4271	&	0.0100	&	20.8869	&	-0.0685	&	-4.4878	&	0.0121	&	42.4	\\
7371.66420	&	34.4457	&	0.0075	&	20.8768	&	-0.2659	&	-4.4614	&	0.0075	&	53.3	\\
7371.67486	&	34.4338	&	0.0070	&	20.9119	&	-0.1224	&	-4.4692	&	0.0069	&	56.4	\\
7371.68506	&	34.4326	&	0.0071	&	20.8996	&	-0.6367	&	-4.4650	&	0.0070	&	55.3	\\
7371.69614	&	34.4302	&	0.0069	&	21.0014	&	-0.0799	&	-4.4731	&	0.0067	&	57.4	\\
7371.70673	&	34.4437	&	0.0064	&	20.9632	&	0.2011	&	-4.4688	&	0.0060	&	60.7	\\
7371.71745	&	34.4300	&	0.0065	&	20.9909	&	-0.1295	&	-4.4773	&	0.0062	&	60.6	\\
7371.72832	&	34.4315	&	0.0065	&	20.9366	&	-0.1761	&	-4.4703	&	0.0064	&	59.7	\\
7371.73898	&	34.4143	&	0.0060	&	20.9340	&	-0.1083	&	-4.4778	&	0.0056	&	64.6	\\
7371.74885	&	34.4342	&	0.0069	&	21.0073	&	-0.1534	&	-4.4696	&	0.0068	&	57.3	\\
7371.76023	&	34.4309	&	0.0092	&	21.0583	&	-0.2884	&	-4.4799	&	0.0108	&	44.6	\\
7371.77033	&	34.4082	&	0.0108	&	21.0351	&	0.0180	&	-4.4715	&	0.0135	&	38.6	\\
7372.67782	&	34.2780	&	0.0548	&	20.2459	&	-0.0747	&	-4.3204	&	0.0816	&	9.3	\\
7372.70837	&	34.4203	&	0.0122	&	20.6497	&	1.3029	&	-4.4660	&	0.0155	&	35.1	\\
7749.58185	&	34.4601	&	0.0088	&	18.9395	&	-0.1852	&	-4.4281	&	0.0086	&	44.7	\\
7749.76345	&	34.4587	&	0.0067	&	18.9701	&	-0.1929	&	-4.4346	&	0.0058	&	59.4	\\
7750.55088	&	34.2998	&	0.0072	&	18.9979	&	0.0767	&	-4.4515	&	0.0069	&	57.0	\\
7750.75882	&	34.3231	&	0.0084	&	18.8521	&	0.1063	&	-4.4493	&	0.0085	&	49.7	\\
7751.70251	&	34.4825	&	0.0083	&	18.9755	&	-0.1709	&	-4.4391	&	0.0083	&	49.8	\\
7754.63363	&	34.3629	&	0.0124	&	21.0214	&	0.1569	&	-4.4369	&	0.0141	&	32.8	\\
7754.74514	&	34.3232	&	0.0122	&	20.9659	&	-0.0801	&	-4.4494	&	0.0148	&	34.6	\\
7755.56052	&	34.4412	&	0.0068	&	20.3272	&	-0.1334	&	-4.4452	&	0.0062	&	58.3	\\
7755.68564	&	34.4804	&	0.0105	&	20.6427	&	-0.1800	&	-4.4695	&	0.0119	&	40.5	\\
7756.54391	&	34.4309	&	0.0144	&	20.9620	&	-0.1692	&	-4.4184	&	0.0174	&	31.2	\\
7756.69497	&	34.4055	&	0.0100	&	21.1398	&	0.5372	&	-4.4221	&	0.0103	&	42.5	\\
7757.56227	&	34.3364	&	0.0062	&	20.5935	&	0.0287	&	-4.4387	&	0.0051	&	64.3	\\
7757.76849	&	34.4096	&	0.0173	&	20.5733	&	-0.0650	&	-4.3885	&	0.0211	&	26.6	\\
7767.63488	&	34.3569	&	0.0135	&	20.6536	&	0.0321	&	-4.4392	&	0.0151	&	32.2	\\
7768.62766	&	34.4216	&	0.0098	&	20.7994	&	-0.0101	&	-4.4454	&	0.0101	&	40.1	\\
7768.68724	&	34.4287	&	0.0074	&	20.8263	&	-0.0352	&	-4.4490	&	0.0067	&	55.8	\\
7769.56123	&	34.4124	&	0.0173	&	20.8606	&	-0.1276	&	-4.4173	&	0.0215	&	26.8	\\
7769.74871	&	34.3985	&	0.0120	&	20.9550	&	0.5301	&	-4.4211	&	0.0134	&	36.7	\\
7770.48664	&	34.3650	&	0.0080	&	20.7612	&	-0.0388	&	-4.4285	&	0.0076	&	51.3	\\
7770.67781	&	34.4047	&	0.0071	&	20.6799	&	-0.0105	&	-4.4350	&	0.0065	&	56.8	\\
7771.46646	&	34.3690	&	0.0067	&	21.0110	&	0.0454	&	-4.4360	&	0.0060	&	60.7	\\
7771.62722	&	34.3601	&	0.0070	&	20.7487	&	0.0950	&	-4.4458	&	0.0063	&	57.5	\\
7772.57081	&	34.4089	&	0.0055	&	20.5872	&	-0.0564	&	-4.4509	&	0.0045	&	71.3	\\
7772.74143	&	34.4355	&	0.0073	&	20.5553	&	-0.4219	&	-4.4504	&	0.0069	&	57.2	\\
7776.46636	&	34.3938	&	0.0094	&	20.8932	&	-0.0770	&	-4.4595	&	0.0129	&	47.7	\\
7776.66380	&	34.4101	&	0.0062	&	20.7195	&	-0.0348	&	-4.4487	&	0.0052	&	65.2	\\
7777.45522	&	34.4037	&	0.0104	&	20.8038	&	-0.0037	&	-4.4347	&	0.0112	&	41.2	\\
7777.68914	&	34.4315	&	0.0116	&	20.5416	&	0.0484	&	-4.4115	&	0.0127	&	37.6	\\
7778.41974	&	34.3745	&	0.0072	&	20.9883	&	0.0391	&	-4.4324	&	0.0066	&	57.8	\\
7778.62073	&	34.3518	&	0.0057	&	20.7787	&	0.1042	&	-4.4338	&	0.0046	&	69.2	\\
7802.45286	&	34.4581	&	0.0068	&	20.4947	&	-0.1972	&	-4.4569	&	0.0063	&	58.7	\\
7802.54483	&	34.4560	&	0.0081	&	20.6486	&	-0.1455	&	-4.4405	&	0.0078	&	50.5	\\
7803.39940	&	34.3822	&	0.0127	&	20.6459	&	-0.0650	&	-4.4355	&	0.0154	&	34.5	\\
7803.54249	&	34.4048	&	0.0107	&	20.4169	&	-0.0496	&	-4.4265	&	0.0117	&	39.2	\\
\hline
 \end{tabular}
\end{table*}

\begin{table*}
\centering
  \contcaption{}  
  \begin{tabular}{cccccccc}
  \hline
Time & RV & $\sigma_{\rm RV}$ & CCF BIS & CCF FWHM & \logr & $\sigma_{\log { R'_{\rm HK}}}$ & S/N  \\
  (BJD$_\mathrm{TDB}$-2,450,000) & (\kms) & (\kms) & (\kms) & (\kms) & & \\
  \hline
7804.40126	&	34.3790	&	0.0078	&	20.8478	&	0.0843	&	-4.4383	&	0.0075	&	52.5	\\
7804.53623	&	34.3812	&	0.0103	&	20.6631	&	-1.4568	&	-4.4316	&	0.0110	&	40.8	\\
7806.44739	&	34.3924	&	0.0087	&	20.5212	&	-0.0588	&	-4.4579	&	0.0088	&	47.8	\\
7807.36241	&	34.3950	&	0.0069	&	21.3158	&	-0.1485	&	-4.4362	&	0.0062	&	59.0	\\
7808.52958	&	34.4034	&	0.0059	&	21.0225	&	-0.0204	&	-4.4413	&	0.0048	&	68.2	\\
7808.63565	&	34.3915	&	0.0082	&	20.9503	&	0.0738	&	-4.4256	&	0.0076	&	51.6	\\
7809.45282	&	34.3884	&	0.0128	&	20.7297	&	-0.2052	&	-4.4406	&	0.0158	&	34.2	\\
7810.54426	&	34.3202	&	0.0132	&	20.3217	&	0.0739	&	-4.4810	&	0.0179	&	33.4	\\
7811.35884	&	34.4775	&	0.0137	&	20.7007	&	-0.2704	&	-4.4359	&	0.0174	&	32.9	\\
7811.52928	&	34.4454	&	0.0136	&	21.1617	&	-0.7146	&	-4.4639	&	0.0176	&	31.8	\\
7814.39794	&	34.3446	&	0.0070	&	21.1399	&	0.2747	&	-4.4598	&	0.0066	&	57.5	\\
7814.60206	&	34.3102	&	0.0073	&	20.8498	&	0.1962	&	-4.4721	&	0.0072	&	57.4	\\
7815.54132	&	34.4647	&	0.0127	&	20.7258	&	-0.3086	&	-4.4604	&	0.0161	&	34.7	\\
7816.42836	&	34.3755	&	0.0074	&	20.5119	&	-0.0049	&	-4.4353	&	0.0067	&	54.8	\\
7816.49993	&	34.3972	&	0.0083	&	20.5529	&	-0.0658	&	-4.4187	&	0.0077	&	50.1	\\
7817.43947	&	34.3840	&	0.0131	&	21.0605	&	0.0611	&	-4.4429	&	0.0160	&	33.9	\\
7833.41663	&	34.4450	&	0.0072	&	21.0347	&	-0.1712	&	-4.4311	&	0.0065	&	56.5	\\
7834.41427	&	34.4005	&	0.0091	&	21.2186	&	-0.0202	&	-4.4272	&	0.0092	&	44.9	\\
7835.44577	&	34.4296	&	0.0094	&	20.8092	&	-0.1197	&	-4.4311	&	0.0095	&	43.6	\\
\hline
  \end{tabular}
\end{table*}

\bsp	
\label{lastpage}
\end{document}